%% file: main.tex
\newif\ifAnonymous
\newif\ifLong
\newif\ifIEEE
\Longtrue
\iffalse
\Longtrue
\documentclass[twocolumns,9pt,runningheads,a4paper]{llncs}
\newcommand{\IEEEauthorblockN}{}
\newcommand{\IEEEauthorblockA}{}
\else
\IEEEtrue
\documentclass[9pt,final,
               conference,compsoc,letterpaper]{IEEEtran}
\newcommand{\subparagraph}{}
\fi

\usepackage[utf8]{inputenc}
\usepackage{array}
\usepackage{comment}
\usepackage{enumitem}
\usepackage{makeidx}
\usepackage{multirow}
\ifIEEE
\iftrue
\newlength{\bibitemsep}
\newlength{\bibparskip}
\let\oldthebibliography\thebibliography
\renewcommand\thebibliography[1]{
  \oldthebibliography{#1}
  \setlength{\parskip}{\bibitemsep}
  \setlength{\itemsep}{\bibparskip}
}
\else
\usepackage{bibspacing}
\fi
\usepackage{titlesec}
\fi
\usepackage{todonotes}
\usepackage{url}
\usepackage{xspace}
\usepackage{color}
\usepackage{listings}
\usepackage[acronym]{glossaries}
\usepackage{cleveref}

\input{preamble/listings.tex}

\input{preamble/glossary.tex}

\graphicspath{{figures/}}

\ifLong
\else
\renewcommand{\subsection}[1]{\noindent\textbf{#1:}}
\renewcommand{\subsubsection}[1]{\noindent\textit{#1:}}
\fi

\begin{document}

\title{Camouflage: Hardware-assisted \acrshort{cfi} \\
for the ARM Linux kernel}
\ifIEEE
\author{
\ifAnonymous
\else
\IEEEauthorblockN{Rémi Denis-Courmont}
\IEEEauthorblockA{Huawei Technologies, Finland\\remi@remlab.net
}
\and
\IEEEauthorblockN{Hans Liljestrand}
\IEEEauthorblockA{Aalto University, Finland\\Huawei Technologies, Finland\\hans@liljestrand.dev}
\and
\IEEEauthorblockN{Carlos Chinea}
\IEEEauthorblockA{Huawei Technologies, Finland\\carlos.chinea.perez@huawei.com}
\and
\IEEEauthorblockN{Jan-Erik Ekberg}
\IEEEauthorblockA{Huawei Technologies, Finland\\Aalto University, Finland\\jan.erik.ekberg@huawei.com}
\fi
}
\else
\titlerunning{Camouflaged Flow Integrity}
\author{Rémi Denis-Courmont \inst{1} \and Hans Liljestrand \inst{2,1} \and Carlos Chinea\inst{1} and Jan-Erik Ekberg\inst{1,2}}
\authorrunning{R. Denis-Courmont et al.}
\institute{Huawei Technologies \and Aalto University}
\pagestyle{headings}
\fi
\date{November 2019}

\newcommand{\aarchsf}{AArch64\xspace}
\maketitle

\begin{abstract}
Software \gls{cfi} solutions have been applied to the Linux kernel for memory protection.
Due to performance costs, deployed software \gls{cfi} solutions are coarse grained.
In this work, we demonstrate a precise hardware-assisted kernel \gls{cfi} running
on widely-used off-the-shelf processors.
Specifically, we use the ARMv8.3 \gls{pauth} extension and present a design that uses it to achieve strong security guarantees with minimal performance penalties.
Furthermore, we show how deployment of such security primitives in the kernel can significantly differ from their user space application.
\end{abstract}

\input{sections/introduction.tex}

\input{sections/background.tex}

\input{sections/threats.tex}

\input{sections/reqs.tex}

\input{sections/design.tex}

\input{sections/implementation.tex}

\input{sections/evaluation.tex}

\input{sections/relatives.tex}

\input{sections/conclusion.tex}

\ifLong
\input{sections/acknowledgements.tex}

\fi

\ifIEEE
\bibliographystyle{IEEEtran}
\bibliography{IEEEabrv,references}
\else
\bibliographystyle{splncs04}
\bibliography{references}
\fi

\ifLong
\appendices
\input{sections/background-extra.tex}

\fi

\end{document}

%% file: preamble/listings.tex
\definecolor{lst:comment}{rgb}{0.0,0.0,0.0}
\definecolor{lst:instructions}{rgb}{0.4,0.1,0.1}
\definecolor{lst:registers}{rgb}{0.1,0.4,0.1}
\definecolor{lst:bg}{rgb}{0.98,0.98,0.98}
\lstdefinelanguage{aarch64}{
    sensitive=false,
    keywords={
        ldp,ldr,stp,str,
        mov,movz,movn,movk,adr,adrp,
        add,sub,eor,and,oor,lsr,lsl,asr,bfc,bfi,bfm,bfxil,sbfx,ubfx,
        b,br,bl,blr,ret,
        cmp,cmn,cbz,cbnz,tbz,tbnz,
        msr,mrs,
        pacia,paciza,pacib,pacizb,pacda,pacdza,pacdb,pacdzb,pacga,
        autia,autiza,autib,autizb,autda,autdza,autdb,autdzb,xpaci,xpacd,
        paciasp,pacibsp,pacia1716,pacib1716,paciaz,pacibz,
        autiasp,autibsp,autia1716,autib1716,autiaz,autibz,
        blraa,blrab,braa,brab,retaa,retab,
    },
    keywordstyle=\bfseries\color{lst:instructions},
    keywords=[2]{
        ip0,ip1,fp,lr,sp,
        w0,w1,w2,w3,w4,w5,w6,w7,w8,w9,w10,w11,w12,w13,w14,w15,
        w16,w17,w18,w19,w20,w21,w22,w23,w24,w25,w26,w27,w28,w29,w30,wzr,
        x0,x1,x2,x3,x4,x5,x6,x7,x8,x9,x10,x11,x12,x13,x14,x15,
        x16,x17,x18,x19,x20,x21,x22,x23,x24,x25,x26,x27,x28,x29,x30,xzr,
    },
    keywordstyle=[2]\color{lst:registers},
    breaklines=true,
    morecomment=[l]{//},
    commentstyle=\itshape\color{lst:comment},
}
\lstdefinestyle{customasm}{
    basicstyle=\footnotesize\ttfamily,
    numberstyle=\footnotesize,
    language=aarch64}
\lstdefinestyle{customc}{language=c,
    basicstyle=\tiny\ttfamily,
    keywordstyle=\bfseries\color{green!40!black},
    commentstyle=\itshape\color{purple!40!black},
    identifierstyle=\color{blue},
    stringstyle=\color{orange},
}

%% file: preamble/glossary.tex
\glsdisablehyper

\newacronym{abi}{ABI}{application binary interface}
\newacronym{aslr}{ASLR}{address space layout randomization}
\newacronym{bti}{BTI}{Branch Target Indicator}
\newacronym{cfg}{CFG}{control flow graph}
\newacronym{cfi}{CFI}{control flow integrity}
\newacronym{dep}{DEP}{data execution prevention}
\newacronym{dfi}{DFI}{data flow integrity}
\newacronym{dop}{DOP}{data-oriented programming}
\newacronym{el}{EL}{exception level}
\newacronym{el0}{EL0}{user mode}
\newacronym{el1}{EL1}{kernel mode}
\newacronym{el2}{EL2}{hypervisor mode}
\newacronym{el3}{EL3}{secure firmware mode}
\newacronym{elf}{ELF}{executable and linkable format}
\newacronym{fdt}{FDT}{flattened device tree}
\newacronym{fp}{\texttt{FP}}{frame pointer}
\newacronym{gcc}{GCC}{GNU Compilers Collection}
\newacronym{gpr}{GPR}{general-purpose register}
\newacronym{hvc}{HVC}{hypervisor call}
\newacronym{ir}{IR}{intermediate representation}
\newacronym{isa}{ISA}{instruction set architecture}
\newacronym{jop}{JOP}{jump-oriented programming}
\newacronym{kasan}{\texttt{KASAN}}{kernel address sanitizer}
\newacronym{kvm}{KVM}{kernel virtual machines}
\newacronym{lkm}{LKM}{loadable kernel module}
\newacronym{lr}{\texttt{LR}}{link register}
\newacronym{lto}{LTO}{linkage time optimization}
\newacronym{mac}{MAC}{message authentication code}
\newacronym{mmu}{MMU}{memory management unit}
\newacronym{oop}{OOP}{object-oriented programming}
\newacronym{os}{OS}{operating system}
\newacronym{pauth}{PAuth}{pointer authentication}
\newacronym{pac}{PAC}{pointer authentication code}
\newacronym{pc}{PC}{program counter}
\newacronym{pic}{PIC}{position-independent code}
\newacronym{prng}{PRNG}{pseudo-random number generator}
\newacronym{pte}{PTE}{page table entry}
\newacronym{rop}{ROP}{return-oriented programming}
\newacronym{simd}{SIMD}{single instruction multiple data}
\newacronym{smc}{SMC}{secure monitor call}
\newacronym{sp}{\texttt{SP}}{stack pointer}
\newacronym{tlb}{TLB}{translation look-aside buffer}
\newacronym{va}{VA}{virtual address}
\newacronym{vla}{VLA}{variable-length array}
\newacronym{vm}{VM}{virtual machine}
\newacronym{vmsa}{VMSAv8}{Virtual Memory System Architecture version 8}
\newacronym{xom}{XOM}{execute-only memory}

%% file: sections/introduction.tex
\section{Introduction}

The Linux kernel open source community has recently directed much effort
on improving the security and attack resistance of the kernel.
This is evident from both a good track record in fixing reported vulnerabilities, but also in the number of new security features introduced in the kernel.
These include security solutions adapted from user space mechanisms such as \gls{aslr}~\cite{kaslr}, $\mathbin{W \oplus X}$ memory policies
or coarse-grained \gls{cfi} integration~\cite{cfi,llvm-cfi}.
Performance costs prevent the use of powerful security solutions such as fine-grained \gls{cfi}.
Nonetheless, the currently integrated mechanisms preclude many crude memory attacks, such as code injection, inside the kernel.

Still, more than 50\% of the recently reported CVEs on kernel vulnerabilities relate to memory errors~\cite{hllkernel}.
Moreover, new advanced attacks, such as, control-flow bending~\cite{cfbending} or \gls{dop}~\cite{dop}, are likely to eventually be used as part of kernel exploit chains.
At the same time, new hardware memory-safety features are introduced
into ARM and Intel processors.
To date, most of the design evolution in this domain has targeted the protection of user applications,
under the assumption that transferring the hardware mechanisms into kernel space is straightforward.
In this paper, we argue that this is only partially true, and use the recent ARMv8.3-\gls{pauth} extension as an example of this.

We apply
\gls{pauth} to the Linux kernel 
to prevent the exploitation of bugs that could, for instance, lead to root privilege escalation or leakage of system secrets.
This requires non-trivial changes and additions to prior \gls{pauth}-based schemes.
Because of programming patterns used withing the kernel,
approaches that focus only on function pointers are ineffective in kernel context.
Our design accommodates and protects such patterns by also protecting pointers to critical data structures, such as the use of \emph{operations} tables that contain function pointers. 
The \gls{pauth} keys are not \emph{banked}, i.e., they must be switched out when entering or exiting kernel space.
We demonstrate a novel design that uses \gls{xom}
to securely set kernel keys without exposing them to an adversary that can read kernel memory.
The contributions of this paper are:
\begin{enumerate}
    \item A secure architecture for kernel \gls{pauth} key management, that does not depend on traps to higher \glspl{el}.
    \item A design for maintaining binary compatibility between non-\acrshort{pauth} processors and a protected kernel and modules.
    \item A \gls{pauth} instrumentation scheme that is compatible and effective with Linux kernel coding patterns.
    \item A hardened \gls{pauth} backwards \gls{cfi} scheme that is robust against replay attacks despite kernel task stack shallowness.
\end{enumerate}

%% file: sections/background.tex
\section{Background}

\subsection{Code reuse attacks on {\aarchsf}}
The \aarchsf call instructions save the function return address in the \gls{lr}.
The canonical prologue for a non-leaf function then stores it and the caller's \gls{fp}
\ifLong
\footnote{unless disabled by compiler optimizations}
\fi
on the stack
(Listing~\ref{lst:frame-record}).
The epilogue conversely restores the \gls{fp} and \gls{lr} values from the stack.
In this scenario, a memory vulnerability---e.g., stack-buffer overflows---may enable an attacker to overwrite the frame record and 
control \gls{lr} 
when the \texttt{RET} is invoked~\cite{smashing}. The 
attacker can then redirect the execution flow to an arbitrary address, e.g., for \gls{rop}~\cite{rop}. Similarly, \gls{jop}~\cite{jop} attacks corrupt instructions pointers in memory before they are consumed
by an indirect jump \texttt{BR} or call \texttt{BLR} instruction.
Alternatively, attacks may target function pointers
indirectly referenced in data structures, such as C++ virtual table pointers~\cite{coop}.

\begin{figure}
    \begin{lstlisting}[showstringspaces=false,style=customasm,
        label={lst:frame-record},caption={
            The \aarchsf frame record is used to store
            and restore the \gls{fp} and \gls{lr} values
            in the function prologue and epilogue, respectively.}
        ]
func:   // Prologue
        stp     fp, lr, [sp, #-16]!
        mov     fp, sp
        // ...
        // Epilogue
        ldp     fp, lr, [sp], #16
        ret
    \end{lstlisting}
\end{figure}

\subsection{\aarchsf \acrfull{pauth}}
ARMv8.3-A introduces the \gls{pauth} extension\ifLong{} (see also \Cref{pauth})\fi{}.
It substitutes unused bits in \aarchsf pointers to store a keyed \gls{mac}.
We follow ARM notation, and call the \gls{mac} a \gls{pac}.
The \gls{pac} is, by default, generated using the QARMA~\cite{qarma} algorithm.
It is derived from a secret key, the pointer's address and a \textit{modifier}.
On load, a pointer can be authenticated against the same secret key and modifier.
\Gls{pauth} can be used to build statistical \gls{cfi} and \gls{dfi}.
Starting from version 5.0, the Linux kernel enables the use of \gls{pauth} in user space
\ifLong
and for \gls{kvm} guests
\fi
but not kernel space.
The LLVM/Clang and GCC compilers support 
backward-edge \gls{cfi} with \gls{pauth}
but use only the \gls{sp} as modifier\ifLong
(Listing~\ref{lst:frame-record-pac}).
\begin{figure}
    \begin{lstlisting}[showstringspaces=false,style=customasm,
    label={lst:frame-record-pac},caption={
        The \acrshort{sp} is signed and authenticated
        in the function prologue and epilogue, respectively.}]
func:   // Prologue
        pacia   lr, sp
        stp     fp, lr, [sp, #-16]!
        mov     fp, sp
        // ...
        // Epilogue
        ldp     fp, lr, [sp], #16
        autia   lr, sp
        ret
    \end{lstlisting}
\end{figure}
If the \gls{lr} value saved on the stack was overwritten
during between the prologue and the epilogue,
then the \texttt{AUTIA} instruction should detect a mismatch in the \gls{pac},
and set \gls{lr} to an invalid address, such that \texttt{RET} triggers
an instruction fault exception, rather than execute the attacker's \gls{rop} chain.

The kernel tracks keys in the per-thread \texttt{thread\_struct} in-kernel structure,
and installs the keys of the \textit{switched-to} thread during user context switch.
The \texttt{exec()} system call will automatically generate a new set of keys
whenever a new user address space is instantiated.
By default, keys are shared by all tasks in a single address space,
but an architecture-specific \texttt{prctl()} call is available
to manually provision keys per thread.
\else{}.
\fi

\subsection{Kernel-user separation}
The Linux kernel implements a 1:1 threading model,
i.e., one kernel task is allocated for each user thread in \gls{el0}.
When a user thread invokes a system call,
execution flow is transferred to the kernel exception handler in \gls{el1}.
\Acrshort{sp} is \textit{banked} on \aarchsf, i.e., 
the processor keeps track of the value separately for each \acrshort{el},
However the \glspl{gpr} and the \textit{system registers}
holding the current \gls{pauth} keys are shared between all \glspl{el}.
Consequently, the exception handler must save the processor state
before executing the system call in \gls{el1},
and afterwards restore it before returning to \gls{el0}.
The same sequence also occurs when user space triggers an exception,
or when an asynchronous interrupt is encountered when a user thread is running.
Because the \gls{pauth} keys are not banked, they must be set on kernel entry if \gls{pauth} is to be used within the kernel.
Conversely, the \gls{pauth} keys of the running user process must be restored when the kernel returns to \gls{el0}.

%% file: sections/threats.tex
\section{Threat model and requirements}
\subsection{Threat model\label{model}}
We assume a powerful adversary with full control over unprivileged user processes, including the capability to launch arbitrary processes and invoke arbitrary system calls.
We further assume that the adversary can use a memory corruption bug in the kernel system call interface to read and write kernel memory.
However, the adversary cannot modify write-protected memory (including \gls{xom}).
This limitation can be realized by locking down \gls{mmu} system control registers and tables via the hypervisor~\cite{rkp}.
Nonetheless, the adversary can leak kernel secrets in readable memory and overwrite pointers in writable memory regions.

%% file: sections/reqs.tex
\newcounter{requirement}
\newcommand{\reqlabel}[1]{\refstepcounter{requirement}\label{req:#1}~(\textbf{R\therequirement})}
\newcommand{\reqref}[1]{~(\textbf{R\ref{req:#1}})}

\subsection{Requirements}
Our goal is to protect kernel call-flow by protecting the integrity of vulnerable pointers.
Specifically, we use \gls{pauth} to fulfill the following requirements:
\begin{itemize}
    \item \textbf{Integrity:} Detect corruption of pointers that affect kernel control flow and mitigate the scope of pointer reuse\reqlabel{integrity}.
    \item \textbf{Robustness:} Protect configuration and confidentiality of \gls{pauth} keys used by the kernel\reqlabel{robustness}.
    \item \textbf{Deployability:} Limit the impact on existing kernel coding patterns and existing kernel code\reqlabel{coding}.
    \item \textbf{Performance:} Minimize performance overhead in terms of execution time and memory use in practical use cases\reqlabel{perf}.
    \item \textbf{Compatibility:} Maintain the existing user space \gls{abi} and \gls{pauth} functionality\reqlabel{abi}.
\end{itemize}

\subsection{Challenges}
The prior kernel support for \gls{pauth} targets only \gls{el0} usage.
When we now apply \gls{pauth} to protect the Linux kernel, we need to address security challenges arising from the different software and hardware context present in \gls{el1}:

\subsubsection{Key allocation}
\Gls{pauth} supports five simultaneously active keys per processor core.
All kernel tasks share the same address space, and so use the same set of keys.
Kernel \gls{pauth} key configuration must be protected to prevent modification of the keys\reqref{robustness}.
However, we must also maintain the existing Linux \gls{abi} on \aarchsf, which guarantees that \gls{pauth} keys are usable in \gls{el0}\reqref{abi}.
Consequently, keys must be changed on kernel entry and exit.

\subsubsection{Key confidentiality}
Kernel keys must remain constant from system boot to system halt so that signed pointers remain verifiable throughout.
Therefore,  when the kernel is not running, the kernel keys must be preserved in \gls{el1} memory so that the \gls{os} scheduler can restore them when switching between \glspl{el}.
To maintain key confidentiality, we must both prevent the reading of the \gls{pauth} keys from system registers and memory\reqref{robustness}.

%% file: sections/design.tex
\section{Design\label{design}}

\subsection{Key management}
\Gls{pauth} keys must be available on boot.
Therefore, the bootloader uses a \gls{prng} to generate them and then stores them in \gls{xom}~\cite{norax} before starting the kernel.
\Gls{xom} is mapped in kernel space
but its permissions are enforced by the hypervisor,
and so it allows secure storage of the keys with minimal performance impact.
The key values are encoded within the executable code of a function that has the sole purpose of writing the kernel keys into the system configuration registers.
Because the code cannot be read it cannot be disassembled to extract the keys.
After completion, the function clears all \glspl{gpr} to prevent the keys from being leaked\reqref{robustness}.
This ensures that kernel keys are usable as soon as the kernel boots.

The kernel only needs to set the \gls{pauth} register values,
it need not be able to read the keys;
hence we can use static code analysis to verify that
no code exists in the kernel, including the \glspl{lkm},
which would read the keys from system registers\reqref{robustness}.
We also check that no code exists
that would corrupt the \gls{pauth} flags in the \texttt{SCTLR\_EL1} register
and thus disable the kernel keys.

\subsection{Backward-edge \acrshort{cfi}}
The reference implementation~\cite{qcom2017} of backward-edge \gls{cfi} with \gls{pauth}
is vulnerable to replay attacks within a given thread because the values of the modifier,
\acrshort{sp}, often repeat.
This problem is accentuated for the task stacks within the kernel, 
where system calls made by a given user thread run with same shallow kernel stack (of 16~KiB).
Moreover, each user thread has its own kernel stack that is aligned on a 4~KiB boundary,
such that the 12 lower order bits of \gls{sp} repeat across threads.
To mitigate such attacks\reqref{integrity}, we construct the modifier by concatenating the low order 32~bits of \gls{sp} with the low order 32~bits of the address of the function, which is inferred from the current \gls{pc}.

\subsection{Pointer Integrity}
Protecting all kernel pointers would be prohibitive for performance reasons~\cite{kdfi}\reqref{perf}.
Instead, we explicitly mark select pointers for protection in the kernel source code.
To ease software engineering efforts, we plan to add a new source code attribute to annotate pointer members in compound type declarations\reqref{coding}.
The compiler could then automatically insert the signing and authenticating \gls{pauth} instructions.
Moreover, this would allow the compiler to use combined \gls{pauth} instructions, such as the \emph{authenticated branch-and-link} \texttt{BLRAB},  instead of a \texttt{PACIB} and \texttt{BLR} pair.
Our evaluated prototype uses inline assembler macros instead, which wrap \gls{pauth} in C code, for use when assigning or evaluating a signed pointer.

To ward off reuse attacks,
we use a modifier constructed by concatenating the low-order 48~bits
of the containing object's address with a 16-bits constant
that uniquely identifies a certain member of a certain object type.
Since \aarchsf uses only 48~bits of address space,
the modifier uniquely identifies the object in memory at a given time.
The 16-bits constant then segregates pointers at the same address based on their type.
The same modifier construction is used to protect function and data pointers\reqref{integrity}.

\subsection{Forward-edge \acrshort{cfi}}
Vulnerabilities stemming from writable and corruptible function pointers within kernel memory are well-known and understood~\cite{kspp}.
Consequently, most kernel function pointers are stored in static \emph{operations} structures.
The operations structures are located in the read-only section \texttt{.rodata}\footnotemark{} which cannot be tampered (\Cref{model}).
\footnotetext{In normal \gls{pic}, the \texttt{.data.relro} section would be used instead.}
This is functionally similar to C++ language virtual method tables,
in the simplified scenario of pure-virtual classes directly inherited
by final concrete derived classes, without multiple inheritance.

Many kernel object types embed pointers to operations structures.
This approach saves memory if more than one object of a given type is allocated,
and mitigates the risk of corrupted function pointers.
Nevertheless, Cook~\cite{cook} notes that forward-edge \gls{cfi} is still necessary for the kernel.
Indeed, there are still writable function pointers in the Linux kernel that need to be integrity protected using \gls{pauth}, including:
1) pointers in hardware-specific device drivers that do not follow best practices, and
2) lone function pointers, which typically are not put in operations structures (since it would not save memory for a single pointer).

\subsection{\Acrfull{dfi}}
Because function pointers are often replaced with operations tables, an attacker could try to modify the pointers to the tables instead of the function pointers.
Consequently, our design must also protect pointers to operations tables.
For instance, the \texttt{struct file} structure describes an open file.
It contains the \texttt{f\_ops} 
pointer to \texttt{const struct file\_operations},
that contains a large number of function pointers.
Those function pointers are provided by the file system or device driver
backing the given open file.
In this case, \texttt{f\_ops} needs to be protected even though it is a data pointer,
not a function pointer,
to ensure effective forward-edge \gls{cfi} in the kernel.
We use the same modifier scheme for both data and function pointers.
We also note that the same approach for protecting pointers could be used to protect other sensitive pointers, such as
the \texttt{f\_cred} pointer to file credentials in the \texttt{struct file} structure.

\ifLong
In conclusion, our full implementation uses 3 of the 5 keys:
\begin{itemize}
    \item one instruction key for backward-edge \gls{cfi},
    \item the other instruction key for forward-edge \gls{cfi},
    \item one of the two data keys for \gls{dfi}.
\end{itemize}
\fi

\subsection{Run-time linkage}
Most kernel pointers are initialized only at run-time,
but a few are set within static/global structure instances.
When such a statically initialized pointer is integrity-protected with \gls{pauth},
its \gls{pac} must be computed before any kernel code attempts to authenticate and use the pointer\reqref{coding}.
For instance, a \texttt{struct work\_struct} object,
describing a deferred execution callback,
can be initialized statically in kernel sources using a C pre-processor macro,
\texttt{DECLARE\_WORK}, instead of using the \texttt{INIT\_WORK} function at run-time.
To address this corner case, a new \gls{elf} section is inserted into the kernel (and \glspl{lkm}).
This section consists of a table of all statically initialized signed pointers, not dissimilar to the existing relocation table sections.
Each entry in the table specifies:
1) the location of a to-be-signed pointer,
2) the \gls{pauth} key to use, and
3) the 16-bit constant for the modifier.
Macros such as \texttt{DECLARE\_WORK} are altered
so that they automatically define and insert an entry in the table.
Then, at early boot, after Linux kernel self-relocation,
the table is iterated through and each pointer is signed in place.
An equivalent procedure is applied when loading an \gls{lkm} at run-time.

%% file: sections/implementation.tex
\section{Implementation}

Our prototype is based on version 5.2 of the Linux kernel, running on QEMU\cite{qemu}.
We believe that our findings broadly apply
to other \textit{Unix}-like kernels on \aarchsf.
We add support for our architecture to a proprietary firmware bootloader and the hypervisor.
This includes the generation of pseudo-random kernel keys at boot time, much like the random seed\footnotemark{} for kernel \gls{aslr}, and \gls{xom}, which is described next.
\footnotetext{passed to the kernel early boot code via the \gls{fdt} on \aarchsf}
Our compiler modifications are based on LLVM 8.0.

\subsection{\Acrfull{xom}}
\begin{figure}
    \centering
    \includegraphics[width=\columnwidth]{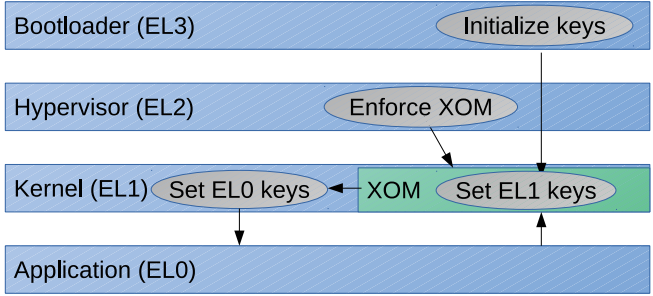}
    \caption{Architecture diagram}
    \label{fig:arch}
\end{figure}

The bootloader generates the pseudo-random \gls{pauth} keys for the kernel and updates the kernel \gls{pauth} key function before the kernel boots (\Cref{fig:arch}).
This conceals the kernel keys without requiring a costly switch to a higher \gls{el} when setting keys at run-time\reqref{perf}.
Each 128-bits \gls{pauth} key is defined by a pair of 64-bits system registers.
The setter runs before interrupts are re-enabled to prevent key leakage and then loads the keys into \glspl{gpr} using the \texttt{MOVZ} and \texttt{MOVK} \emph{move-immediate} instructions that encode the values in the instructions themselves.
The keys are then assigned from the \glspl{gpr} with \texttt{MSR}.
All relevant \glspl{gpr} are zeroed out before the function returns.
The memory page containing the function is mapped as \gls{xom} by the hypervisor, which prevents:
1) reading the immediate values from the instructions,
2) writing to modify the code or keys, and 
3) execution in \gls{el0}.
As shown in~\cite{rkp}, \gls{xom} security properties are enforced by a proprietary hypervisor, which also prevents, for instance, tampering with \gls{mmu} system registers.

\subsection{Return address protection}
To contain the risk of a replay attack against backward-edge \gls{cfi},
we change the \gls{pauth} modifier used for signing return addresses,
so that it varies by the called function.
The compiler is modified to emit function prologues and epilogues
as in listing~\ref{lst:prologue-bfi}.
\begin{figure}
    \begin{lstlisting}[showstringspaces=false,style=customasm,
        label={lst:prologue-bfi},caption={
            The \acrshort{sp} is signed with a custom modifier.}]
function: // Prologue: sign LR
        adr     ip0, function
        mov     ip1, sp
        bfi     ip0, ip1, #32, #32
        pacib   lr, ip0
        stp     fp, lr, [sp, #16]!
    \end{lstlisting}
\end{figure}
The move-from-SP instruction (Line 3) is necessary because \aarchsf does not allow
\gls{sp} as an operand of a bit~field move instruction (Line 4).
LLVM-generated \aarchsf code only saves and restores \gls{sp} from memory
when a variable size stack allocation occurs,
and as of Linux version 5.0
\gls{sp} is always restored by adding an immediate value to it (as on line 6)
so it cannot be corrupted.

We also provide functionally equivalent prologue and epilogue patterns in assembler
macros \texttt{frame\_push} and \texttt{frame\_pop}.
These need to be used in hand-written functions,
such as in optimized \gls{simd} procedures,
but also in the context-switching function \texttt{cpu\_switch\_to}.
In that particular function,
we additionally need to sign the switched-from kernel task's \gls{sp}
and authenticate the switched-to task's \gls{sp} using our pointer integrity scheme,
so as to protect the \glspl{sp} of tasks that are scheduled out.

\subsection{Pointer Integrity\label{pi}}
Like outlined in \Cref{design}, most kernel function pointers are read-only and need no memory protection. However, 
a number of them do remain, predominantly within specific device drivers.
A semantic search using Coccinelle~\cite{coccinelle}
over the complete Linux version 5.2 source code yields 1285 function pointer members assigned at run-time, residing in 504 different compound types.
We expect that for 229 out of the 504 types---i.e., those with more than one function pointer---should follow existing kernel practices~\cite{kspp}
and be converted to use read-only  \textit{operations} structures.

Our prototype employs a unified approach to protect pointers, i.e.,
1) the remaining isolated function pointers, 
2) sensitive, writable data pointers, and 
3) data pointers to read-only \textit{operations} structures.
We use a \gls{pauth} modifier with a 16-bits constant identifying the combination of the containing type and the compound type member,
combined with the 48-bits address of the containing object.
We define in-line \gls{pauth} assembler convenience macros
to easily sign and authenticate the pointers.
For instance, to access or assign the \textit{operations} pointer of an open file:
\begin{itemize}
    \item One setter, \texttt{set\_file\_ops()},
        signs and stores the pointer into a \texttt{struct~file} object, e.g.: \\
        \texttt{
            const struct file\_operations my\_ops = ... \\
            /* ... */ \\
            struct file *fp; \\
            /* ... */ \\
            set\_file\_ops(fp, \&my\_ops);
        }
    \item One getter, \texttt{file\_ops()} loads,
        authenticates and returns the operations pointer from an object
        (Listing \ref{lst:call-bfi}), e.g.: \\
        \texttt{
            struct file *fp; \\
            /* ... */ \\
            file\_ops(fp)->read(fp, buf, len, NULL);
        }
\end{itemize}
\begin{figure}
    \begin{lstlisting}[showstringspaces=false,style=customasm,
        label={lst:call-bfi},caption={
            A file \textit{operations} pointer is authenticated before an indirect call.}
        ]
        // load signed fp->f_ops from fp (x0)
        ldr     x8, [x0, #40]
        mov     w9, #0xfb45
        bfi     x9, x0, #16, #48 // modifier
        autdb   x8, x9 // authenticate f_ops
        ldr     x8, [x8, #16] // load read
        blr     x8 // call read pointer
    \end{lstlisting}
\end{figure}

Based on these patterns,
we have written a Coccinelle~\cite{coccinelle} \textit{semantic patch} that can semi-automatically adjust the kernel source code whenever a structure member is used in the kernel.
Specifically we substitute the direct reading and writing of protected pointers
with explicit \texttt{get} and \texttt{set} inline functions.
Then the \texttt{get} and \texttt{set} functions are manually patched
to invoke \gls{pauth} instructions.

\subsection{Brute force mitigation}
\Glspl{pac} can have up to 31~bits, but with typical Linux page and \acrlong{va} configurations the space remaining for the \glspl{pac} is 15 bits\ifLong{} (see \Cref{vmsa})\fi{}.
This lies well within practical reach of a brute force attack by an attacker-controlled local application. Consecutive pointer authentication failures must therefore be limited.

When the authentication of a pointer fails,
a memory fault exception is raised due to an invalid memory address.
By default, a memory fault inside the Linux kernel will unconditionally terminate the user process (\texttt{SIGKILL} signal),
and depending on the context also trigger an \textit{OOPS} and halt the system.
We change the kernel configuration to halt after a limited number of \gls{pauth} failures have occurred, as they constitute a strong indication of an attempt at kernel bug exploitation.

\subsection{Backward compatibility}
Our implementation has a build-time option to issue machine code for either \aarchsf version~8.3 (or higher) only, or for older versions.
To support this, \gls{pauth} includes backwards-compatible
\texttt{PACIB1716} and \texttt{AUTIB1716} instructions, which behave as no-ops on older processors.
As no such instructions exist for \gls{pauth} data (D) keys, in this case we use the same key for instruction- and data pointer protection.

%% file: sections/evaluation.tex
\section{Evaluation}

\subsection{Performance evaluation}
For functional verification, we use the \aarchsf system emulation from the open-source QEMU\cite{qemu} tool, which supports ARMv8.3-A and \gls{pauth}. 
However, QEMU is not cycle-accurate and hence it cannot be used for performance evaluation.
As hardware with ARMv8.3-A is not yet generally available, we replace all \gls{pauth} instructions with the PA-analogue used in prior work on \gls{pauth}~\cite{parts}.
The PA analogue is an instruction sequence that exhibits the estimated computational overhead of \gls{pauth}, i.e., 4-cycles per instruction.
In the same spirit, all writes to \gls{pauth} key system registers (which do not exist on ARMv8.0-A) are substituted with writes to another register without side effects,  namely \texttt{CONTEXTIDR\_EL1}. 
For performance evaluation, we run our customized Linux kernel on a RaspBerry Pi 3 ARMv8-A device.
We compare our measurements against a baseline measured on the Ubuntu v5.0 kernel running on the same hardware.
All error bars show standard deviation for $n = 20$.

\subsubsection{Key management}
We measured an overhead for switching between kernel and user mode \gls{pauth} keys, upon system call or user mode interrupt, of 9 cycles per key
(measurement average: $8.88$; variance: $.004$).
In our micro-benchmarks, we use three different keys: one data key for \gls{dfi} and two instruction keys for forward- and backward-edge \gls{cfi}, respectively.

\subsubsection{Return address protection}
\begin{figure}
    \centering
    \includegraphics[width=\columnwidth]{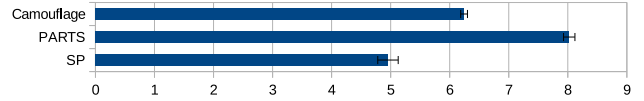}
    \caption{\label{fig:becfi-modifiers}Function call overhead ({nanoseconds}): 1) Proposed solution:32 bits \acrshort{sp} + function address), 2) PARTS: 16 bits \acrshort{sp} + 48 bits LLVM LTO func.id., 3) \acrshort{sp} as  supported by Clang.}
\end{figure}

Backward-edge \gls{cfi} adds a fixed overhead per function call, except for functions optimized to omit their stack frame.
\Cref{fig:becfi-modifiers}
compares the run-time cost of 3 different approaches for computing the \gls{pauth} modifier. We note that our proposal is slightly slower than the weaker protection present in compilers, but faster than prior work with equal security properties for return address protection~\cite{parts}.

\subsubsection{System calls}
\begin{figure}
    \centering
    \includegraphics[width=\columnwidth]{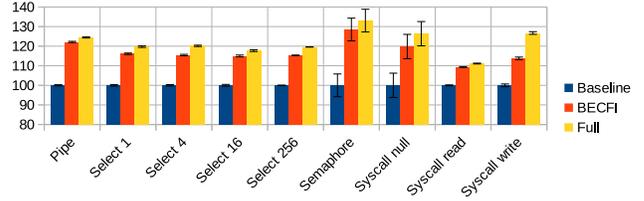}
    \caption{lmbench (relative) latencies with full, backward-edge \acrshort{cfi} and no protection}
    \label{fig:lmbench-lat}
\end{figure}

\begin{figure}
    \centering
    \includegraphics[width=\columnwidth]{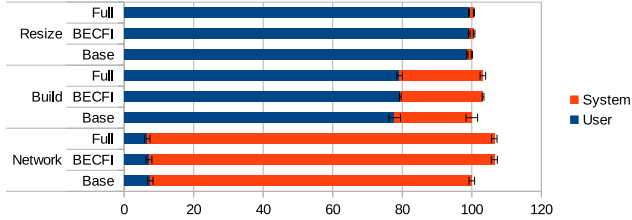}
    \caption{User-space performance: 1) JPEG picture resize (predominantly user computation), 2) Debian package build (balanced), 3) Network download (mostly kernel). Presented with full, backwards-edge CFI and no instrumentation}
    \label{fig:user-loads}
\end{figure}

The performance impact at system call level is measurable as double-digit percentual overhead, as measured using the \texttt{lmbench}~\cite{lmbench} kernel micro-benchmarking tool (\Cref{fig:lmbench-lat}).
The impact is due to a comparatively high rate of function calls to computation, as is visible in kernel system call implementations.
When measured with user space workloads (\Cref{fig:user-loads}),
the geometric mean of the overhead drops to less than $4\%$.

\subsection{Security Evaluation}
For our design to protect kernel control flow\reqref{integrity}
we must protect the integrity of all necessary pointers
and prevent the leakage of \gls{pauth} keys\reqref{robustness}.
As discussed in \Cref{design},
these pointers are function pointers in writable memory and data pointers
to the \textit{operations} structures.
Based on our threat model (\Cref{model}), read-only memory cannot be corrupted as the memory mappings are protected by the hypervisor.
Other protected pointers are signed and verified using \gls{pauth}.

\subsubsection{\Acrshort{pauth} reuse attacks}
We use \gls{pauth} to protect three types of pointers:
1) function return addresses,
2) writable function pointers, and
3) data pointers to \textit{operations} tables.
In all cases, \gls{pauth} detects the injection of arbitrary unsigned pointers.
The attacker is thus forced to either guess---with a probability of $2^{-\textrm{pac\_size}}$---or perform a reuse attack.
Return addresses are protected with a hardened modifier scheme that combines the \texttt{SP} with the function address.
The function address does not completely prevent reuse, but significantly reduces the scope of attacks. 
For code and data pointers, the modifier is tied to the pointer type and its storage location (\Cref{design}).
As most of the \gls{pauth}-protected pointers are long-lived and assigned only once, this effectively prevents most reuse attacks\reqref{integrity}.
An attack is only possible, when a pointer is replaced with another pointer of the same type.
This could happen either because an assigned pointer is legitimately replaced or because dynamically allocated structure is assigned the same address as a previously de-allocated one.

\subsubsection{Key confidentiality}
When stored in memory, the confidentiality of the \gls{pauth} keys is provided by \gls{xom}.
During execution, the keys are loaded into registers from \gls{xom} and could thus be leaked if the key-setting process were preempted.
To avoid this, the key-setting function disables preemption and clears all registers before returning.
The \gls{pauth} keys are readable in the configurations registers.
However, because \texttt{MRS} system register read instructions immediately address the read register,
key reads can be trivially found and rejected (e.g., when loading a module).

\subsubsection{Kernel \acrshort{pauth} verification oracles}
Our threat model includes an adversary with arbitrary user space access.
Consequently the attacker could try to find a \gls{pauth} verification oracle.
The user space process uses a randomly assigned key, and thus cannot verify kernel pointers.
Our design introduces a threshold for allowed \gls{pac} authentication failures to prevent non-critical sections of kernel code from being used as an oracle.
Any failures are also logged, ensuring that such vulnerable code paths can be fixed.

\subsection{Compliance}
Our solution does not retain ISO C language~\cite{c18} semantics, since we bind the object / function address to the \glspl{pac},
e.g., functions like \texttt{memcpy()} or byte-wise pointer copying / casting does fail without code adaptation.
Also, extra assumptions made by the Linux kernel, such as:
1)~All pointer types have the same representation or
2)~Null pointer values are represented by zero bits~\cite{elf-aarch64}, do not hold for \gls{pauth}.
Our protection does provide, e.g., strong protection against replay,
which would not be possible if those assumptions were met.
We conclude that the benefit of strong memory protection outweighs the lack of compliance.

\subsection{Approach analysis}
We select protected pointers for \gls{dfi} and forward-edge \gls{cfi}
only in a semi-automated manner.
For the Linux kernel, this brings finer grained coverage and lower run-time overhead,
considering the large amount of read-only pointers, which need no protection.
This solution also provides better replay protection
than would be permitted by pure ISO C language~\cite{c18} rules.
We surmise that protecting all writable function pointers can be achieved
with reasonable engineering effort within the Linux community.

%% file: sections/relatives.tex
\section{Related work}
Most of the work on hardware-assisted memory protection with ARM \gls{pauth}
has had user applications as their focus. First out was the Qualcomm white-paper~\cite{qcom2017} 
on simple return-edge \gls{cfi} with \gls{pauth},
using \gls{sp} as the only modifier.
Apple proposal~\cite{2pac2furious,applellvm} for forward-edge protection is similar to ours. However, their approach to protect \textit{vtable}'s differs:
the \textit{vtable} pointers are protected with zero as \gls{pauth} modifier.
This approach preserves \texttt{memcpy}, but is susceptible to reuse attacks.
Also, Apple signs all pointers in the \textit{vtable}.
In contrast, we store them in read-only memory to avoid unnecessary signing of pointers.

The first academic presentation on the subject was PARTS~\cite{parts},
who presented a more fine-grained solution
for both forward- and backward-edge \gls{cfi} protection, as well as data pointer protection.
For backward-edge \gls{cfi}, our hardened solution offers security equal to PARTS.
However, to assign unique function identifiers, PARTS requires \gls{lto}, which is intrinsically incompatible with \glspl{lkm}
and not (yet) widely supported by Linux build systems.
We improve on the modifier construction as discussed in \Cref{pi}.
Furthermore, we also remedy a PARTS shortcoming,
where the lower 16~bits of \gls{sp} are prone to replay attacks
\emph{across} different threads, whose respective stacks may well be separated
by an exact multiple of $2^{16} = 65536$~bytes within the kernel address 
space.

The \gls{pauth} mechanism has also been applied to stack canaries~\cite{pcan},
and to a chained (authentication) solution~\cite{pacstack}
for fully protecting the stack frames against reuse attacks.
None of these projects considered the kernel as a target for their protection efforts,
and with the exception of the authenticated stack,
they are not readily applicable for kernel protection.

Ferri et al\cite{ferri2019towards} proposed managing (application) keys
in \gls{el2} (or \gls{el3}) using dedicated traps to a higher \gls{el}.
Kernel keys could potentially also be managed that way.
However the hypervisor traps that this scheme relies on
exist primarily to prevent \acrlongpl{vm} from accessing \gls{pauth} system registers
under a legacy \gls{pauth}-unaware hypervisor.
The traps can also be used for lazy initialization of \gls{pauth} support by the hypervisor,
but they are not intended and optimized for frequent occurrence.

For memory protection in the Linux kernel, PaX~\cite{pax2015} implements
backward- and forward-edge \gls{cfi}  in software
respectively using stack canaries and type-based cookie matching.
Moreira \emph{et al.}~\cite{kcfi} refines software forward-edge kernel \gls{cfi}
by disallowing indirect calls to never indirectly referenced functions.
Recent patches\footnotemark{} by ARM enable return address protection
but expose the kernel \gls{pauth} keys in memory.
\footnotetext{\url{https://marc.info/?l=linux-arm-kernel&m=157416679432588}}

%% file: sections/conclusion.tex
\section{Conclusions and Future work}

To date, there has been a lack of academic contributions on how to apply hardware-assisted memory protection features in processors to the operating system kernel.
In this work, we provide a first architectural glimpse on how the Linux kernel can accommodate pointer integrity.
Moreover, we show that this can be done at acceptable performance cost by using hardware assistance such as \gls{pauth}.
We also convey the insight that memory and memory references are treated differently in the kernel, compared to how they are handled in user space. 
This hopefully also encourages more research on OS kernel memory protection; today there are several memory-protection mechanisms that have appeared or have been announced for both ARM and Intel platforms, and all of them could by themselves or in combination be used to strengthen the \gls{os} kernel against run-time attacks. 

For this particular work, there are also a few unexplored directions that we leave for future examination.
Attacks targeting the interrupt handler could potentially modify or replace kernel register content, e.g. modifiers or previously authenticated pointers.
Register spills pose a similar threat to the integrity of register content; indicating that their protection would further reduce the attack surface.
Last, this work does not consider the next obvious step with authenticated pointers: an integrity-protected kernel system call \gls{abi}
where kernel and user space protection can maintain \gls{pauth} security guarantees across  across privilege boundaries. 
However, we note that the \gls{pauth} extension could be extended to further support the kernel protection setting.
At the \gls{isa} level, an extension could support layered key management such that the hypervisor can manage the kernel keys without the need for \gls{xom}.
To allow a hardened \gls{abi} with cross-layer signed pointers,  this might also require a processor flag to select the active--i.e., kernel or user---set of keys.

In conclusion, our work demonstrates how to realize call-flow protection within the kernel by leveraging the ARMv8.3-A \gls{pauth} extension.
Our evaluation indicates that such comprehensive protection can be achieved with minimal performance overhead (less than $<4\%$ \texttt{lmbench}).
By accounting for practical deployment limitations, e.g., performance constraints and preservation of existing \glspl{abi}, this work is applicable beyond a research setting.
This work demonstrates that although the kernel imposes such restrictions, it also offers opportunities for novel kernel-specific adaptions of existing security mechanisms.

%% file: sections/acknowledgements.tex
\section*{Acknowledgements}
This work was supported by the European Research Institute of Huawei Technologies.

%% file: sections/background-extra.tex
\section{\Acrfull{vmsa}} \label{vmsa}

The \aarchsf \gls{vmsa} represents
virtual memory addresses, or pointers, as 64-bits values.
However the virtual address space does not use the entirety of the 64~bits.
The maximum virtual address space size is 48~bits -- or 52~bits
with the Large Virtual Address extension, ARMv8.2-LVA.
Additionally, \gls{vmsa} divides the address space in two ranges,
each with their own translation tables.

For a given virtual memory address, the translation table is selected
by bit 55 (noted $x$ below).
By convention, the first translation table (\texttt{TTBR0\_EL1}) maps
the address space of the current user process,
while the second table (\texttt{TTBR1\_EL1}) maps kernel addresses.
In a typical run-time configuration,
such as that of Ubuntu operating system,
the address space has 49 bits:
In addition to bit 55, the lower $49 - 1 = 48$~bits are used.
The remaining bits are sign-extended (\Cref{tab:vmsa-ranges}).

\begin{table}
 \caption{\label{tab:vmsa-ranges}\Acrshort{vmsa} address ranges}
 \begin{tabular}{|l|r|l|}
 \hline
 Address range & Bit 55 & Usage \\
 \hline
 $\mathtt{0xffffffffffffffff}-\mathtt{0xffff000000000000}$ &
 $1$ & Kernel \\
 \hline
 $\mathtt{0xffffefffffffffff}-\mathtt{0x0001000000000000}$ &
 & Invalid \\
 \hline
 $\mathtt{0x0000ffffffffffff}-\mathtt{0x0000000000000000}$ &
 $0$ & User \\
 \hline
 \end{tabular}
\end{table}

\subsection{Address tagging}
\gls{vmsa} optionally ignores the top byte (bits 56-63) of addresses.
Linux enables this feature for user space addresses but leaves it
disabled for kernel-space addresses (except in debug builds
with the \gls{kasan} enabled).
Assuming the usual page size of 4~KiB, user and kernel addresses
are laid out as shown in \Cref{tab:vmsa-pointer}, where $t$ are ignored (tag) bits,
and $a$ are addressing bits.
8~bits and 16~bits are effectively meaningless sign extension respectively for user and kernel addresses.

\begin{table}
 \caption{\label{tab:vmsa-pointer}\aarchsf pointer on Linux}
 \centering
 \begin{tabular}{| c | c | c | c | c |}
  \hline
  \multicolumn{5}{| c |}{User pointer ($x = 0$)} \\
  \hline
  Tag & $x$ & Sign extension & Page number & Page offset \\
  \hline
  63-56 & 55 & 54-48 & 47-12 & 11-0 \\
  \hline
  $tttttttt$ &
  $0$ &
  $000...000$ &
  $aaa...aaa$ &
  $aaa...aaa$ \\
  \hline
  \hline
  \multicolumn{5}{| c |}{Kernel pointer ($x = 1$)} \\
  \hline
  Sign ext. & $x$ & Sign extension & Page number & Page offset \\
  \hline
  63-56 & 55 & 54-48 & 47-12 & 11-0 \\
  \hline
  $11111111$ &
  $1$ &
  $111...111$ &
  $aaa...aaa$ &
  $aaa...aaa$ \\
  \hline
 \end{tabular}
\end{table}

\subsection{\Acrfull{xom} on \aarchsf}
\Gls{xom} is a type of memory mapping which has only execute permission,
i.e., neither read nor write permission.
\Gls{xom} is already available with \gls{vmsa} for user space applications (in \gls{el0}).
However the translation table format of \gls{vmsa} is such
that any memory mapping is implicitly readable at \gls{el1},
which precludes \gls{xom} in kernel.

To achieve \gls{xom} in \gls{el1}, we need to use the second stage of translation,
which comes with \aarchsf hardware virtualization (\gls{el2}).
In that case, the read permission can be controlled in the translation table
by the hypervisor.

\section{\aarchsf \acrlong{pauth}} \label{pauth}

ARMv8.3-A, the third major revision of ARMv8-A,
introduces the \gls{pauth}.
This extension adds three new classes of machine instructions to \aarchsf:
\begin{itemize}
\item \texttt{PAC}\ldots instructions sign a pointer:
they replace the "unused" bits with a \gls{mac}
before storing the authenticated pointer to memory.
\item \texttt{AUT}\ldots instructions authenticate a pointer:
they check that the \gls{mac} matches the pointer value
after loading a 64-bits authenticated pointer it from memory.
\item \texttt{XPAC}\ldots instructions strip the \gls{mac}
from an authenticated pointer.
This is primarily intended for debugging purposes.
\end{itemize}

The \gls{mac} is based on an implementation-defined (i.e. processor-dependent)
cryptographic hash algorithm with a 128-bits secret key, 64-bits input
and 32-bits output.
In reference implementations, QARMA~\cite{qarma} is used as the hash algorithm,
although this is not mandated by ARM~\cite{armv8a}.

The sign extension bits are substituted with bits from \gls{mac},
forming the \gls{pac};
extraneous \gls{mac} bits are discarded.
An authenticated pointer would match the following pattern,
where $c$ represents a \gls{mac} bit.

\subsection{Keys}

\gls{pauth} supports five distinct keys simultaneously on a given processor core:
\begin{itemize}
\item two keys, IA and IB, to sign instruction pointers,
i.e. function pointers or function call return addresses,
\item two keys, DA and DB, to sign data pointers, and
\item one key, GA, to sign generic data separately
(not constrained by the address space layout).
\end{itemize}

Each key is configured at run-time via dedicated privileged system registers.
Since a system register can only hold 64 bits, two registers are defined for
each key, or ten registers in total.

Also each key has its own signing instruction:
\texttt{PACIA}, \texttt{PACIB}, \texttt{PACDA}, \texttt{PACDB}
and \texttt{PACGA}.
Likewise, each key except for the generic data key,
has its own authenticating instruction:
\texttt{AUTIA}, \texttt{AUTIB}, \texttt{AUTDA} and \texttt{AUTDB}.

\subsection{Modifier}

The different keys allow for a limited degree of segregation
between different contexts.
For instance, out of the two instruction keys:
\begin{itemize}
\item On the one hand, key IA could be used to sign return addresses,
providing backward-edge \gls{cfi}.
\item On the other hand, key IB could be used to sign function pointers,
providing forward-edge \gls{cfi}.
\end{itemize}

To protect against trivial replay attacks,
whereby two pointers signed the same key,
a cryptographic salt is necessary.
That is the role of the "modifier" register.